\documentclass[letterpaper,11pt]{article}
\usepackage{jheppub}
\usepackage{amsmath}
\usepackage{bm}
\usepackage{amssymb}
\usepackage{listings}
\usepackage{multirow}
\usepackage{makecell}
\usepackage{array}
\usepackage{subfigure}
\usepackage{siunitx}
\usepackage{booktabs}
\usepackage{amssymb}
\usepackage{pifont}
\usepackage{listings}
\usepackage{xcolor}
\usepackage{float}
\usepackage{slashed}

\usepackage{tikz}
\usepackage{pgffor}							

\newcommand{\be}{\begin{equation}}
\newcommand{\ee}{\end{equation}}
\newcommand{\beq}{\begin{equation}}
\newcommand{\eeq}{\end{equation}}
\newcommand{\bea}{\begin{eqnarray}}
\newcommand{\eea}{\end{eqnarray}}
\newcommand{\besp}{\begin{equation}\begin{split}}
\newcommand{\eesp}{\end{split}\end{equation}}

\newcommand{\Dfbd}{\mathord{\buildrel{\lower3pt\hbox{$\scriptscriptstyle\leftrightarrow$}}\over {D}_{\mu}}}

\def\0{\textbf{0}}
\def\1{\textbf{1}}
\def\2{\textbf{2}}
\def\3{\textbf{3}}
\def\4{\textbf{4}}
\def\5{\textbf{5}}
\def\6{\textbf{6}}
\def\7{\textbf{7}}
\def\8{\textbf{8}}
\def\9{\textbf{9}}

\begin{document}

\title{Analysis of p-Wave Sommerfeld Resonance Behavior for Finite-Size Dark Matter}

\author[a]{Wu-Long Xu}
\author[b,c]{,~Jin Min Yang}
\author[d,e]{,~Rui Zhu}

\affiliation[a]{School of Integrated Circuits, Chengdu Technological University, Chengdu 611730, P. R. China}
\affiliation[b]{Center for Theoretical Physics, Henan Normal University, Xinxiang 453007, P. R. China}
\affiliation[c]{Institute of Theoretical Physics, Chinese Academy of Sciences, Beijing 100190, P. R. China}
\affiliation[d]{State Key Laboratory of Dark Matter Physics, Tsung-Dao Lee Institute, Shanghai Jiao Tong University, Shanghai 200240,  P. R. China }
\affiliation[e]{Key Laboratory for Particle Astrophysics and Cosmology (MOE) \& Shanghai Key Laboratory for Particle Physics and Cosmology, Shanghai Jiao Tong University, Shanghai 200240, P. R. China}

\emailAdd{396440567@qq.com(corresponding author)}
\emailAdd{jmyang@itp.ac.cn}
\emailAdd{zhu.rui@sjtu.edu.cn(corresponding author)}

\abstract{
Different stages of cosmic evolution impose different requirements on the dark matter annihilation cross section. Sommerfeld enhancement provides a possible way to accommodate them. However, in the point-like dark matter scenario, satisfying these requirements with either $s$-wave or $p$-wave Sommerfeld enhancement can strongly constrain the dark matter and mediator masses and their coupling. A finite dark matter size introduces an additional parameter and may relax these constraints. Motivated by this possibility, in this work we study the $p$-wave Sommerfeld resonances of finite-size dark matter  (like proton or neutron) without specifying its internal structure. We find that the $p$-wave contribution can exceed the $s$-wave contribution. Nevertheless, the finite-size $p$-wave resonances are weaker than those in the point-like case. Despite this suppression, their velocity dependence remains similar to that of point-like dark matter. We then extend our analysis to nugget dark matter composed of a small number of constituents. In this case, the $p$-wave Sommerfeld enhancement exhibits behavior similar to that of point-like dark matter.
}

\maketitle


\section{Introduction}\label{sec1}
Dark matter remains one of the most mysterious components of the Universe. Its microscopic properties, including its mass, spin, and size, remain largely unknown~\cite{Clowe:2006eq,Cirelli:2024ssz}. Moreover, direct-detection experiments have found no conclusive evidence for dark matter so far. Taken together, these facts motivate us to consider an alternative possibility~\cite{Buchmueller:2017qhf,Jungman:1995df,XENON:2025vwd}: present-day dark matter may not consist of elementary point-like particles~\cite{Lin:2019uvt,Monteiro:2020wcb}, instead, it may be composed of spatially extended states formed during cosmic evolution. Such states could resemble protons, glueballs, or even dark atoms~\cite{Cline:2021itd,Sannino:2026hkc,Carenza:2024qaq}.

Studies of finite-size dark matter usually begin by assuming a specific internal structure. Typical examples include dark atoms and dark nugget~\cite{Bai:2018dxf,Gresham:2017zqi,Gresham:2017cvl,Wise:2014jva,Wise:2014ola,Grabowska:2018lnd,Belli:2026woj}. The resulting scattering or annihilation cross sections are then used to constrain parameters such as the dark matter mass and size~\cite{Bringmann:2016din}. By contrast, recent studies have proposed a more phenomenological approach. Instead of specifying a microscopic model, finite-size dark matter can be characterized by an internal charge distribution~\cite{Chu:2018faw}. Once this distribution is specified, both the dark matter self-scattering cross section and its scattering cross section with nucleons can be calculated~\cite{Xu:2026fsj,Wang:2023wbw,Xu:2020qjk}. For example, finite-size effects can generate a velocity-dependent self-scattering cross section. They therefore provide an alternative to light-mediator scenarios and may help alleviate the small-scale structure problems of cosmology~\cite{Wang:2023wrx,Wang:2021tjf}. Moreover, the same finite-size effects can lead to distinct regimes in dark matter–nucleon scattering, including the Born, resonant, and classical regimes~\cite{Xu:2025xaq}. Consequently, meaningful constraints may be placed on the dark matter mass, size, and interaction strength without specifying its detailed microscopic structure.

In addition to scattering, dark matter annihilation provides another important probe of dark matter physics. Most existing studies focus on point-like dark matter~\cite{Ding:2021zzg}. However, the annihilation rate is tested in cosmological and astrophysical environments with markedly different characteristic velocities. Around recombination, the dark matter velocity can be as low as $v\sim10^{-8}c$, although its precise value depends on the thermal history. Energy injection from dark matter annihilation can modify the cosmic microwave background (CMB) anisotropies~\cite{Masi:2015fca}. The Planck 2018 data impose the constraint $p_{\rm ann}\equiv
f_{\rm eff}\langle\sigma v\rangle/m_\chi
<3.2\times10^{-28}\,
\mathrm{cm^3\,s^{-1}\,GeV^{-1}}$
at the 95\% confidence level~\cite{Planck:2018vyg}. In dwarf spheroidal galaxies, the characteristic velocity is $v\sim10^{-4}c$. Gamma-ray observations therefore probe the annihilation cross section at a higher velocity. For TeV-scale dark matter annihilating into $\tau^+\tau^-$, the upper limit is typically of the order of $10^{-25}$--$10^{-24}\,\mathrm{cm^3\,s^{-1}}$~\cite{Fermi-LAT:2025gei}. The characteristic velocity in the Galactic halo today is even higher, with $v\sim10^{-3}c$. Depending on the annihilation channel and the observational data set, the relevant cross section can be around $10^{-24}\,\mathrm{cm^3\,s^{-1}}$~\cite{Prantzos:2010wi,Daylan:2014rsa}. The precise value also depends on the dark matter mass and the assumed halo profile. Finally, at thermal freeze-out, the characteristic velocity is $v\sim0.3c$. The observed relic abundance conventionally requires $\langle\sigma v\rangle_{\rm fo}
\simeq 2\text{--}3\times10^{-26}\,
\mathrm{cm^3\,s^{-1}}$~\cite{Feng:2010zp,Arkani-Hamed:2008hhe}. A velocity-dependent Sommerfeld enhancement may help reconcile the annihilation rates probed in these different environments~\cite{Hartl:2025mtb}. 

The annihilation cross section can be expanded in partial waves. After including the Sommerfeld effect, the cross section takes the schematic form
$\sigma v=aS_0(v)+b v^2S_1(v)+\cdots$,  where the first and second terms correspond to $s$-wave and $p$-wave annihilation, respectively~\cite{Sommerfeld:1931qaf,Watanabe:2025kgw,An:2012uu,Parikh:2024mwa}. For point-like dark matter, the Sommerfeld-enhanced $s$-wave rate generally increases as the velocity decreases, before eventually reaching saturation. Consequently, a large annihilation rate in the Galactic halo can lead to tension with constraints from dwarf spheroidal galaxies and the CMB. By contrast, Sommerfeld-enhanced $p$-wave annihilation can exhibit a nonmonotonic velocity dependence~\cite{Kamada:2023iol,Beneke:2024iev,Pradhan:2026maz}. Its rate may be enhanced at intermediate velocities while remaining suppressed at very low velocities~\cite{Iengo:2009ni,Das:2016ced}. The CMB constraint can therefore be substantially weakened. Nevertheless, the viable parameter space may still be narrow~\cite{Wang:2025tdx}. In particular, it depends sensitively on the dark matter mass, the mediator mass, and their coupling.

As noted above, finite-size effects provide another source of velocity dependence in dark matter self-scattering~\cite{Wang:2023xii}. It is therefore natural to ask whether they can also modify Sommerfeld-enhanced annihilation. Such a modification may offer a new way to reconcile the annihilation rates probed at different velocities. In our previous work, we investigated the size dependence of the $s$-wave Sommerfeld enhancement in detail~\cite{Xu:2026fsj}. In this work, we extend the analysis to $p$-wave annihilation. We will first study the resonant behavior of the $p$-wave Sommerfeld enhancement for finite-size dark matter. Then we will consider nugget dark matter composed of smaller constituents with a specified internal structure. Our results provide a theoretical basis for future constraints on dark matter models from annihilation cross sections.

This work is organized as follows. In Sec.\ref{sec2} we investigate how the inclusion of a size parameter affects the $p$-wave Sommerfeld enhancement. In Sec.\ref{sec3} we characterize the $p$-wave Sommerfeld resonances of finite-size dark matter using two parameters. We also explain why the resonance boundaries remain well defined while the structure inside them becomes blurred. Then, we study the $p$-wave resonance behavior of nugget dark matter composed of smaller constituents. Sec. \ref{sec4} gives our conclusions.

\section{Size effects on the p-wave Sommerfeld enhancement}\label{sec2}

In this section, we mainly aim to show the effect of finite size on the $p$-wave Sommerfeld enhancement of dark matter annihilation. First, we consider finite-size dark matter without an internal structure and specify only its charge-density distribution. To illustrate the physics, we choose a simple top-hat profile, namely, $\rho(r)=\frac{3}{4\pi R_{\chi}^3}\theta(R_{\chi}-r)$,
where, $R_{\chi}$ is the radius of dark matter and $r$ is the relative distance between dark matter particles. Therefore, by integrating the charge over the entire volume in coordinate space, the potential $V(r)$ between two finite-size dark matter particles can be obtained as  (the detailed calculation can be found in Appendix~A of Ref.~\cite{Wang:2023xii})
\begin{align}\label{eq1}
V(r) & = \begin{cases}
~- g(r,y) & r<2R_{\chi}\, , \\
\hspace{5cm}\  & \ \\[-6.mm]
~-\alpha\frac{e^{-m_{\phi}r}}{r} 
\times h\left(y\right)&  r>2R_{\chi}\,,
\end{cases}
\end{align}
where $y=R_{\chi}m_{\phi}$ is  the ratio between the radius of a size dark matter particle and the force range among these two particles,  and the dark fine structure constant $\alpha = g_{\chi}^2/4\pi$ with $g_{\chi}$ being the coupling constant. For completeness, the explicit expressions for $g(r,y)$ and $h(y)$ are given in Appendix~\ref{appa}.

Next, the Sommerfeld enhancement factor can be expressed as ~\cite{Iengo:2009ni}
\be \label{eq2}
S=\left( \frac{1\cdot 3\cdots (2l+1)}{C} \right)^2,
\ee
where $l=0$ for the $s$-wave and $l=1$ for the $p$-wave. To obtain the constant $C$, we need to solve the radial Schrödinger equation for the reduced two-body wave function $R_{p,l}$. The corresponding potential is given in Eq.~(\ref{eq1}). The numerical procedure for solving this equation was described in detail in our previous work~\cite{Xu:2026fsj}. That work studied the $s$-wave Sommerfeld enhancement of finite-size dark matter. Therefore, we do not repeat the procedure here. In this work, we focus on the contribution of a higher partial wave, i.e., the $p$-wave, to the Sommerfeld enhancement factor. We also analyze its resonant behavior.
\begin{figure}[ht]
	\centering
	\includegraphics[width=7.5cm]{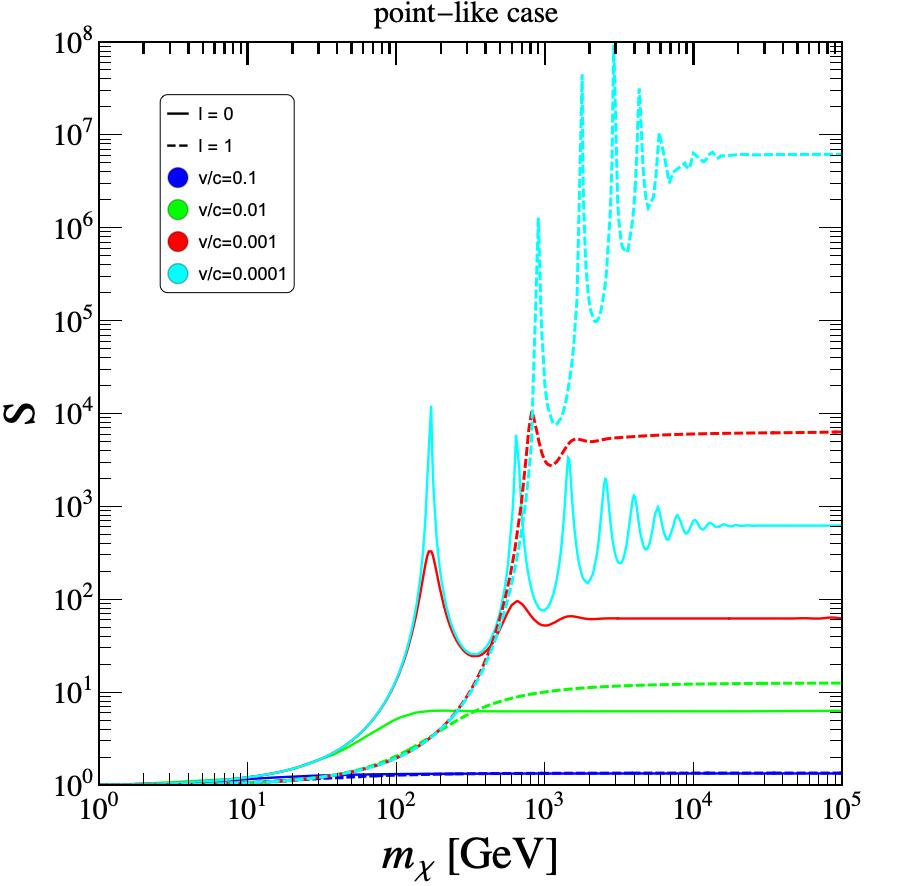}		
	\includegraphics[width=7.5cm]{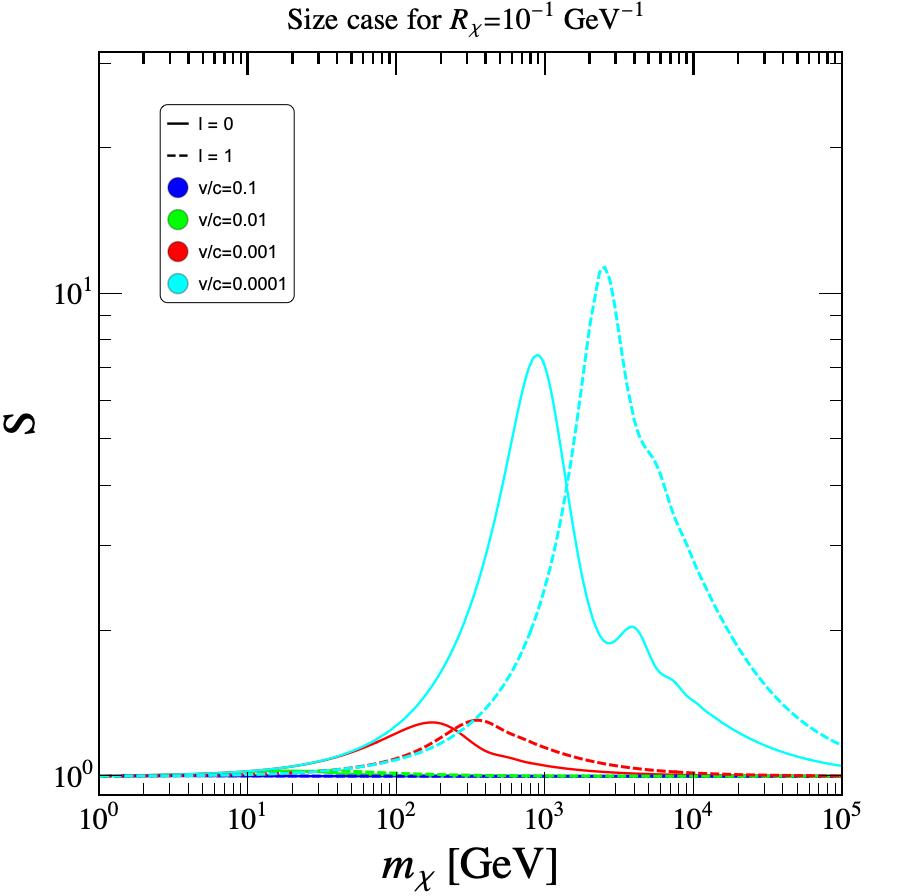}\\	
	\includegraphics[width=7.5cm]{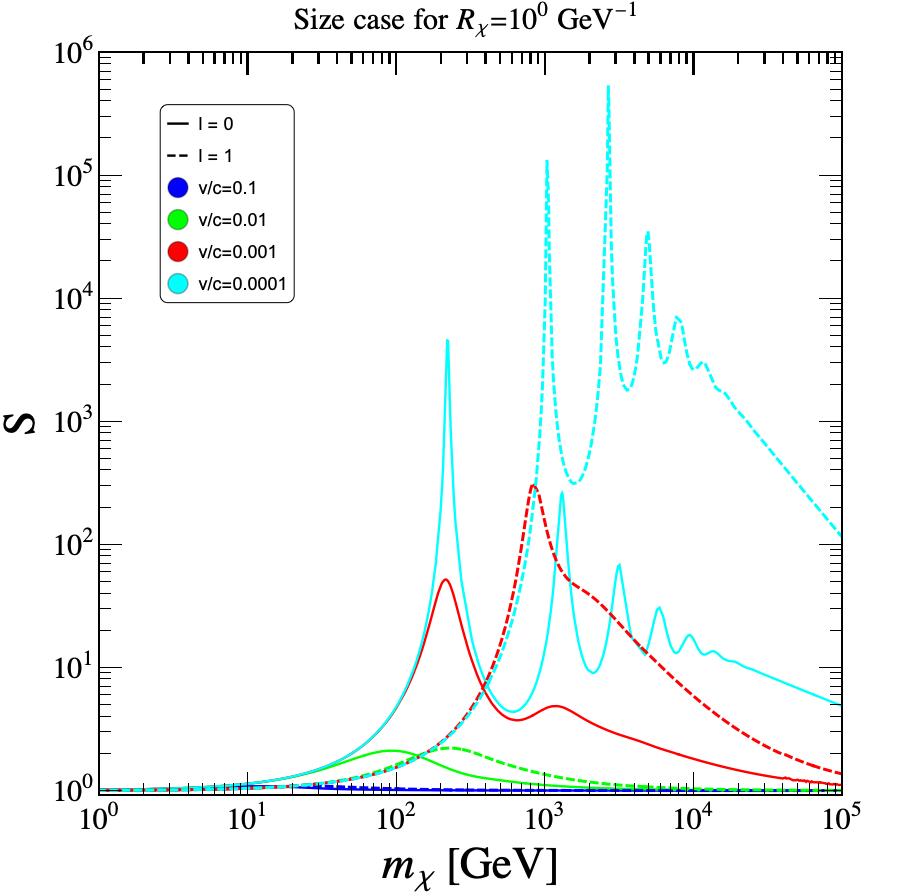}			
	\includegraphics[width=7.5cm]{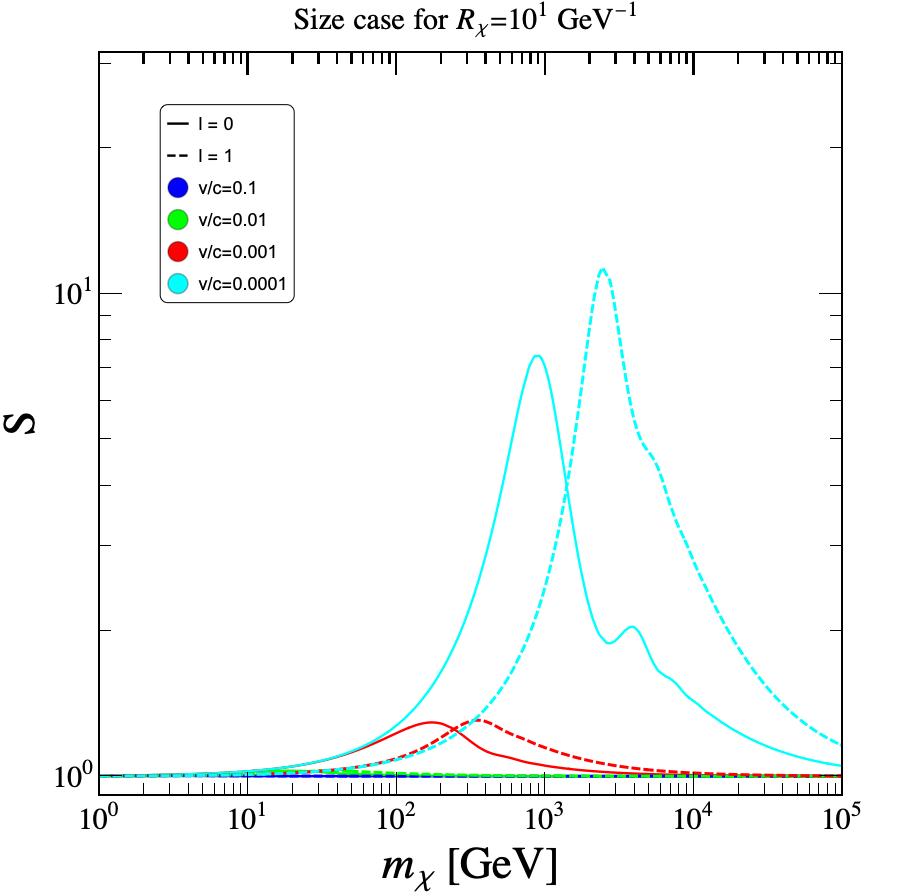}\\	
	\caption{The $s$-wave and $p$-wave sommerfeld enhancement factor  for the Yukawa and puffy potentials with $m_{\phi}=1~\rm GeV$ and $\alpha=0.01$ at different velocities:  $v/c=0.1$(blue color), $v/c=0.01$(green color), $v/c=0.001$(red color) and $v/c=0.0001$(cyan color).
	}
	\label{fig1}
\end{figure}

For comparison with the Sommerfeld resonances of point-like particles, we also numerically calculate the $s$-wave and $p$-wave Sommerfeld enhancement factors for the Yukawa potential. The results are shown in the upper-left panel of Fig.~\ref{fig1}. We then use this panel as a reference. The other panels show the effects of different dark matter sizes at different velocities. Both the $s$-wave and $p$-wave resonance peaks decrease once the particles acquire a finite size. This can be seen from the upper-left panel, where the resonance peaks reach their largest values. Therefore, finite-size effects weaken the $p$-wave resonances. We obtained a similar result for the $s$-wave in our previous work.  We further compare the $s$-wave and $p$-wave resonances for dark matter with different sizes. The dashed and solid curves with the same color correspond to the same dark matter size. The $p$-wave resonances are clearly stronger than the $s$-wave resonances. This result suggests that higher partial waves may dominate the Sommerfeld resonances of finite-size dark matter.

\begin{figure}[ht]
	\centering
	\includegraphics[width=7.5cm]{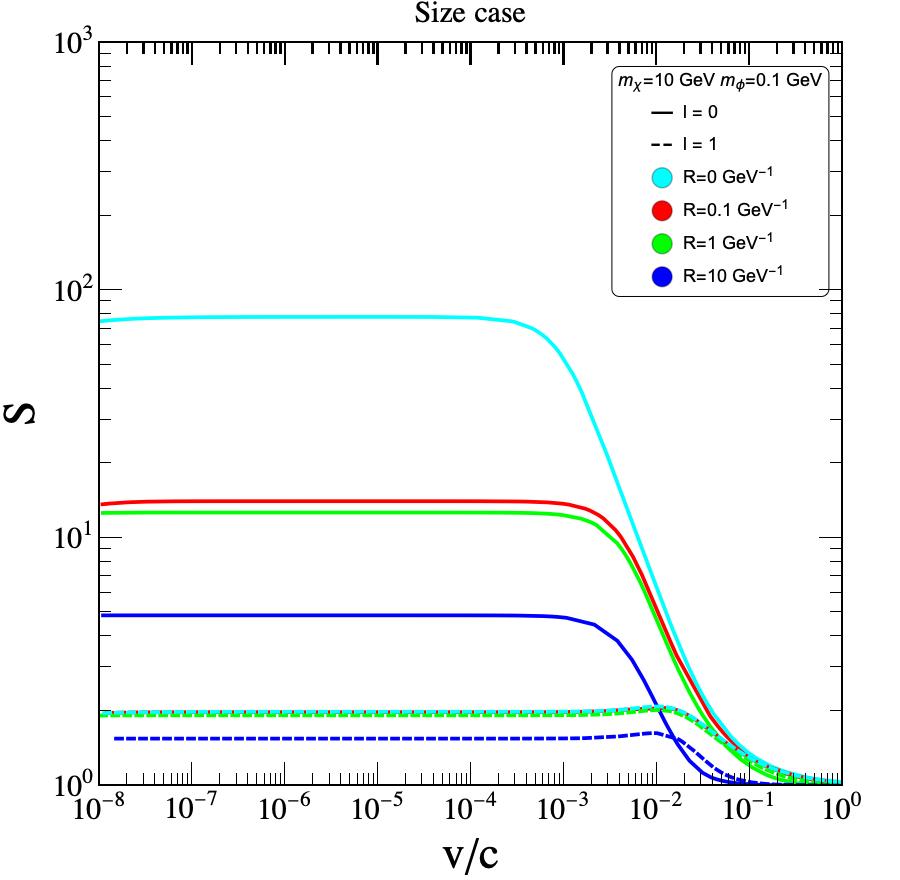}		
	\includegraphics[width=7.5cm]{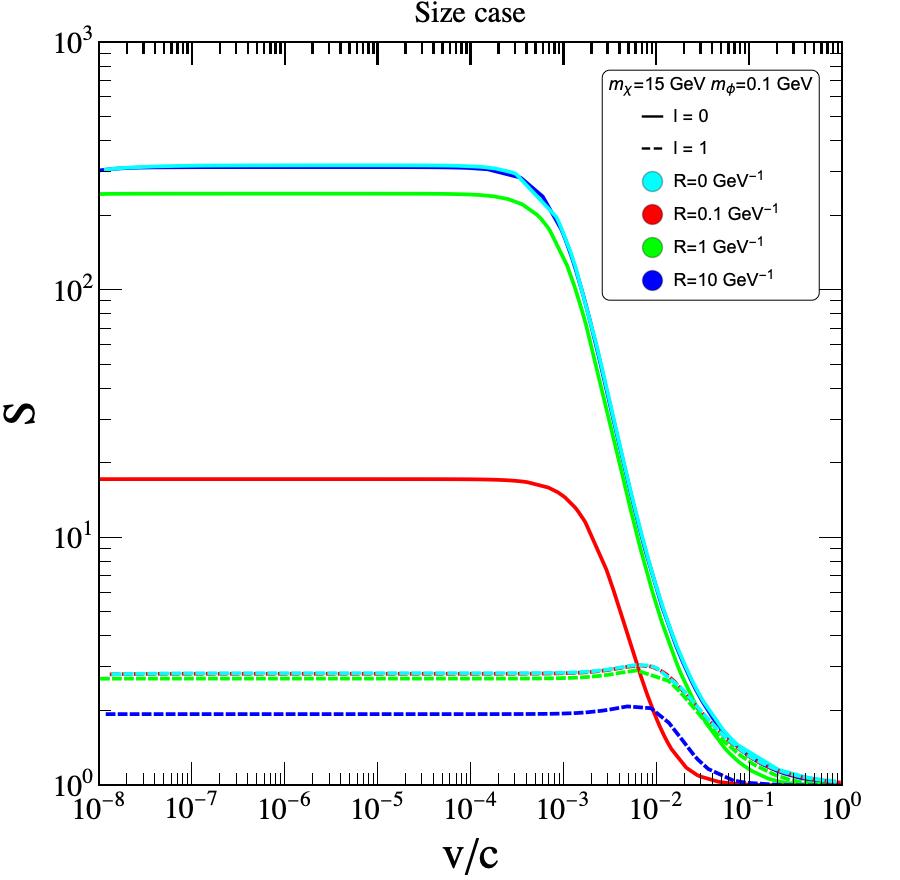}		
	\includegraphics[width=7.5cm]{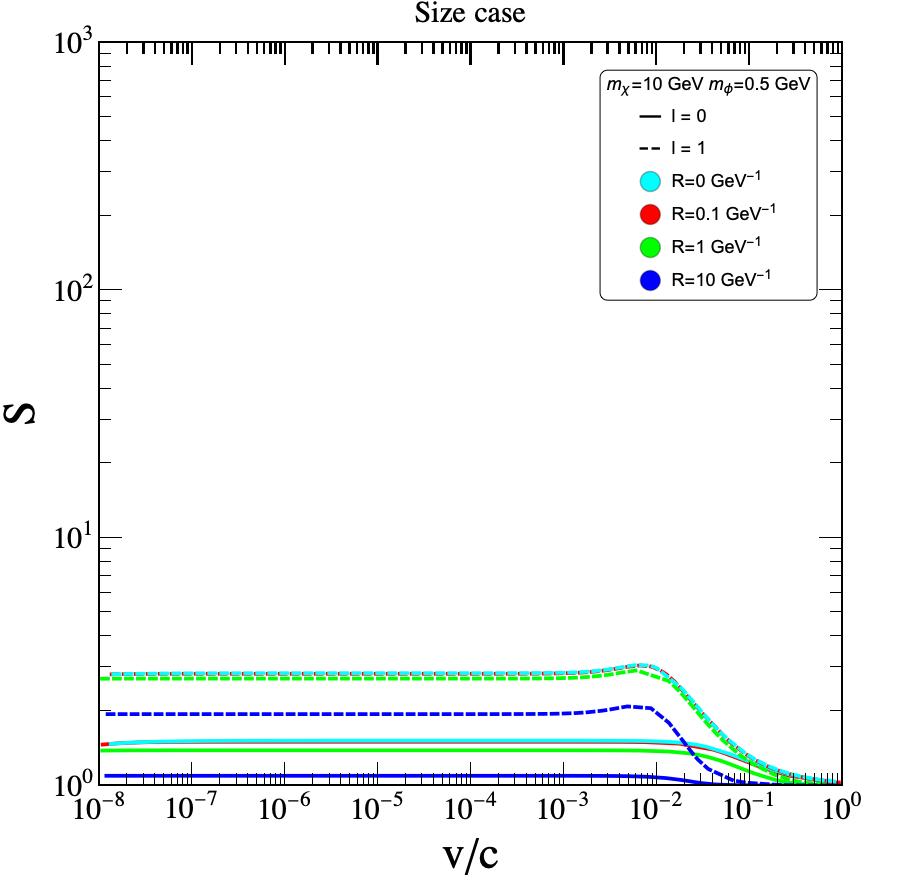}\\	
	\caption{Velocity- dependence of the Sommerfeld enhancement factor for different dark matter radii, dark matter total masses $m_{\chi}$, and mediator masses $m_{\phi}$, with $\alpha=0.01$.
	}
	\label{fig2}
\end{figure}

To address the different annihilation cross sections at different velocities, we also need to analyze the velocity dependence induced by finite-size effects. In Fig.~\ref{fig2}, the blue-green solid and dashed curves represent the $s$-wave and $p$-wave Sommerfeld enhancement factors, respectively. They are compared with the solid and dashed curves of other colors. After finite-size effects are included, the velocity dependence of both the $s$-wave and $p$-wave enhancement factors remains almost unchanged. Each panel of Fig.~\ref{fig2} also shows a difference between the $s$-wave and $p$-wave Sommerfeld resonances. As the velocity decreases, the $s$-wave Sommerfeld enhancement increases monotonically and then approaches saturation. The $p$-wave Sommerfeld enhancement shows a slightly different behavior. Near saturation, it develops a small bump that first rises and then falls. In addition to the $v^{-2}$ factor affecting the total annihilation rate near the saturated $p$-wave resonance, this small nonmonotonic feature may also contribute to the bell-shaped profile of the total annihilation rate.
We also compare the $s$-wave and $p$-wave enhancement factors for different radii in the three panels of Fig.~\ref{fig2}. The blue solid and dashed curves correspond to larger radii. Their Sommerfeld enhancement factors are clearly smaller. This indicates that a larger dark matter size may weaken the Sommerfeld enhancement under the same conditions. An exception can occur near a resonance. For example, in the middle panel of Fig.~\ref{fig2}, the blue $s$-wave curve for a radius of $10~\mathrm{GeV}^{-1}$ is clearly higher than the red curve for a radius of $0.1~\mathrm{GeV}^{-1}$.

Therefore, we got to know that the $p$-wave Sommerfeld resonances are always stronger than the $s$-wave resonances. This conclusion applies to both finite-size and point-like dark matter. Although finite-size effects reduce the Sommerfeld enhancement factor, they leave its velocity dependence almost unchanged. This behavior differs from that of the dark matter self-scattering cross section associated with the same bound states. In the self-scattering case, finite size itself provides an additional source of velocity dependence besides the light mediator. Having studied in detail how finite-size effects modify the $p$-wave Sommerfeld enhancement, we next analyze $p$-wave resonant behavior with a finite interaction size.

\section{Analysis of p-wave resonance behavior}\label{sec3}

For the $p$-wave Sommerfeld resonances of finite-size dark matter, we aim to determine the locations of the zero-energy resonant states. Namely, we study the positions and heights of the resonance peaks. First, we need to identify a parameter that characterizes the resonance positions. For point-like dark matter in the $p$-wave case, such a parameter is $b=\alpha m_{\chi}/m_{\phi}$. In the finite-size case, our previous works~\cite{Xu:2026fsj,Wang:2023wbw} provide some useful clues. In particular, the resonant behavior can be displayed in the plane of two dimensionless parameters, $(bf(y),a\sqrt{y}/f(y))$. Here, $a=v/\alpha$, and the expression for $bf(y)$ is given below:
\small
\begin{eqnarray}
bf(y)
= \frac{ \alpha m_{\chi}}{m_{\phi}}
\frac{3(-15+y^2(15-10y+4y^3)+15(1+y^2)e^{-2y})}{10y^6}\,,
\label{pufb}
\end{eqnarray}
\normalsize
where  function $f(y)$ is defined for the estimation of the validity~\footnote{For point-like dark matter, the parameter $b$ is chosen according to the condition for the Born approximation, $m_\chi\left|\int_{0}^{\infty}rV_{\rm Yukawa}(r)\,dr\right|=b<1$.
By contrast, the system enters the nonperturbative regime when $b>1$. Similarly, for finite-size dark matter, we use the corresponding Born-approximation condition for the finite-size potential, $m_\chi\left|\int_{0}^{\infty}rV_{\rm size}(r)\,dr\right|=bf(y)<1$. This condition was obtained in Ref.~\cite{Wang:2023wbw}.}. For finite-size dark matter, we first consider a model specified only by its charge-density distribution. We do not specify its detailed internal structure. This is the case discussed in the previous section. Protons and neutrons provide examples of such finite-size particles. In addition, some finite-size particle models have explicit internal structures. Examples include dark atoms and dark matter nuggets. Therefore, in the following, we analyze the $p$-wave Sommerfeld resonances in two cases. The first is generic finite-size dark matter, and the second is nugget-type dark matter.

\subsection{Finite-size dark matter case}\label{sec3-1}
\begin{figure}[ht]
	\centering
	\includegraphics[width=6.8cm]{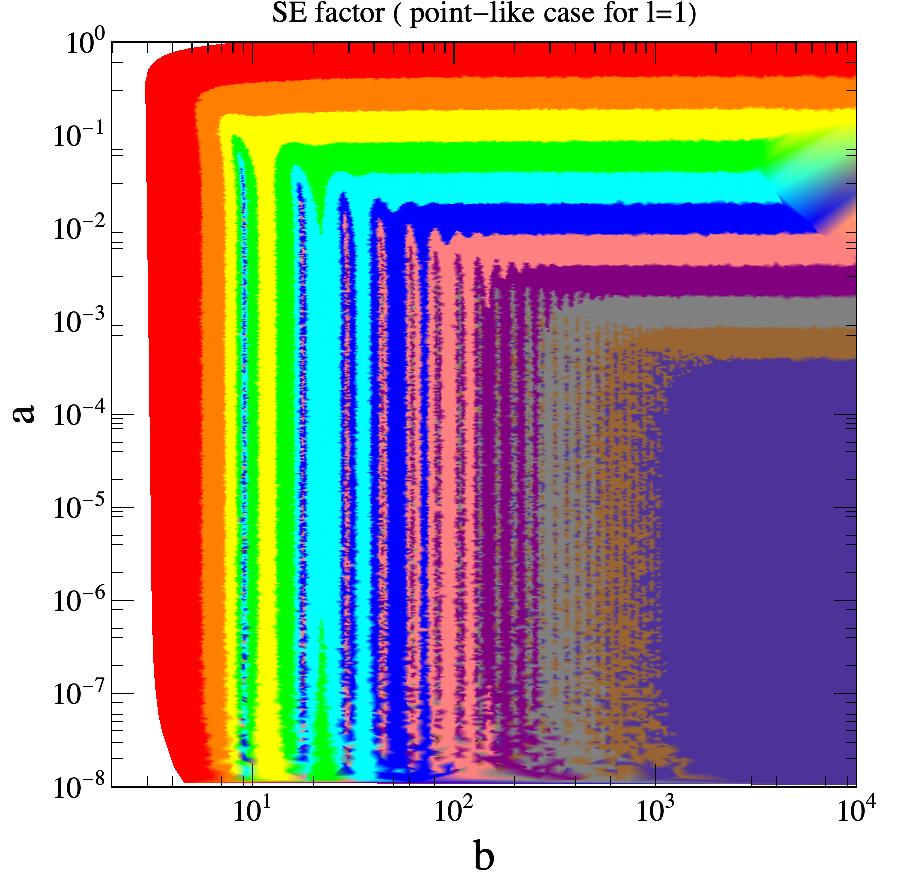}
	\includegraphics[width=8.05cm]{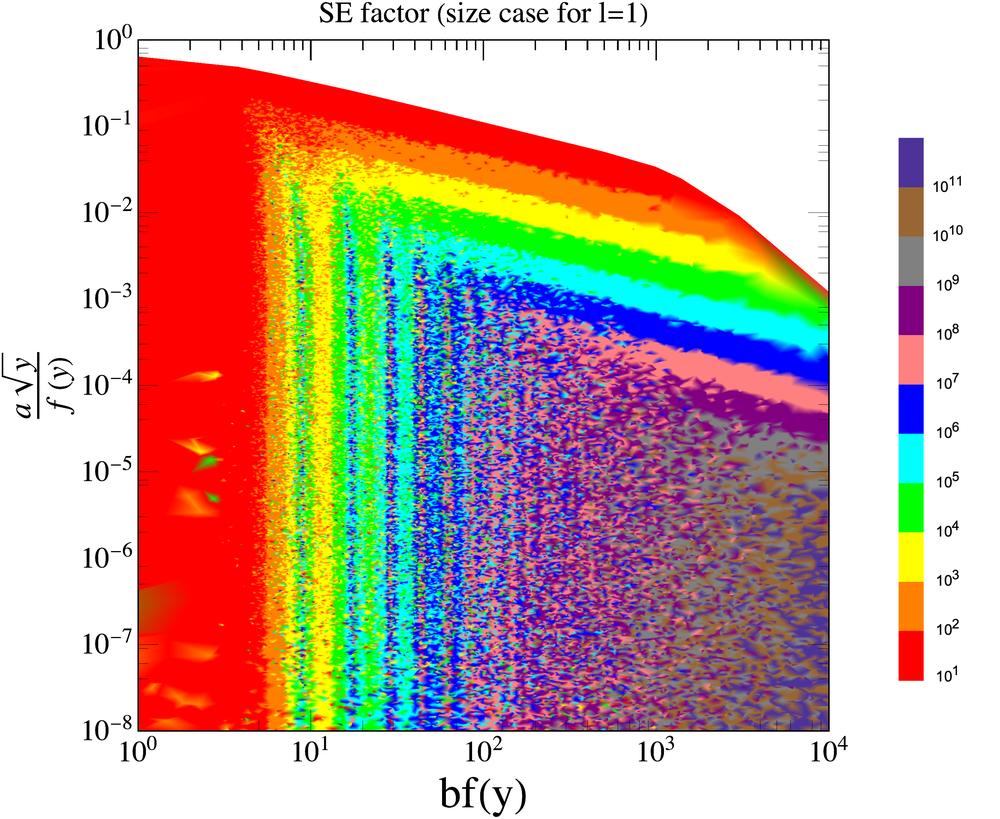}\\	
	\includegraphics[width=7.5cm]{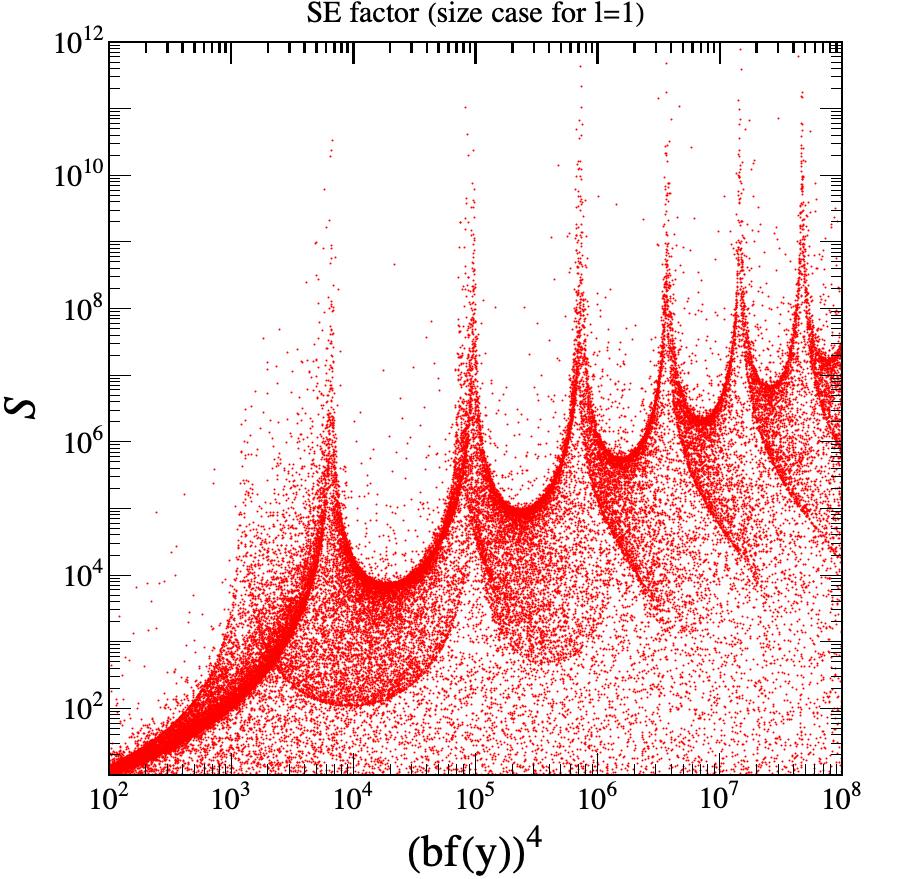}\\	
	\caption{Up panel: The p-wave sommerfeld enhancement factor in the $\sqrt{y}a/f(y)$ versus $b f(y)$ plane for puffy dark matter annihilation (right) and ($a, b$)  for point-like case (left); Bottom panel: The p-wave sommerfeld enhancement factor $S$ as a function of $(b f(y))^4$ where the fourth power is chosen to make the width of the resonance more visible.
	}
	\label{fig3}
\end{figure} 

For the analysis of $p$-wave Sommerfeld resonances, we first consider finite-size dark matter without specifying its detailed internal structure. The model is characterized only by a charge-density distribution. The Sommerfeld enhancement factor can then be obtained by solving the Schrödinger equation for the corresponding potential. For comparison, we first present the $p$-wave Sommerfeld resonances of point-like dark matter in the $(a,b)$ parameter space. The corresponding contour plot is shown in the upper-left panel of Fig.~\ref{fig3}. The resonance positions can be clearly identified. For example, the first zero-energy resonant state occurs at $b=9.082$. Moreover, the $p$-wave contribution to the Sommerfeld enhancement can be very large. In particular, the enhancement factor can reach $S>10^{11}$. The upper-right panel of Fig.~\ref{fig3} shows the $p$-wave Sommerfeld resonances of finite-size dark matter. Here, we use the parameter space $(bf(y),a\sqrt{y}/f(y))$. By comparison, the Sommerfeld enhancement is weaker in the finite-size case. Near $a\sqrt{y}/f(y)=1$, the enhancement factor is smaller than 10. This suppression becomes more evident at larger values of $bf(y)$. In addition, the resonance boundaries of finite-size dark matter remain relatively clear. However, the internal structure of each resonance region becomes blurred. It is less distinct than that in the point-like case. To clarify the origin of this behavior, we enlarge the parameter $bf(y)$ to $[bf(y)]^4$. The results are shown in the lower panels of Fig.~\ref{fig3}. We find that the finite-size $p$-wave resonances are similar to the $s$-wave resonances found in our previous work. Each resonance is no longer represented by a single peak. Instead, it extends over a finite width. We consider this behavior to be natural. When the dark matter size is a free parameter, the resonance peak should vary with the size.

\begin{figure}[ht]
	\centering
	\includegraphics[width=7.65cm]{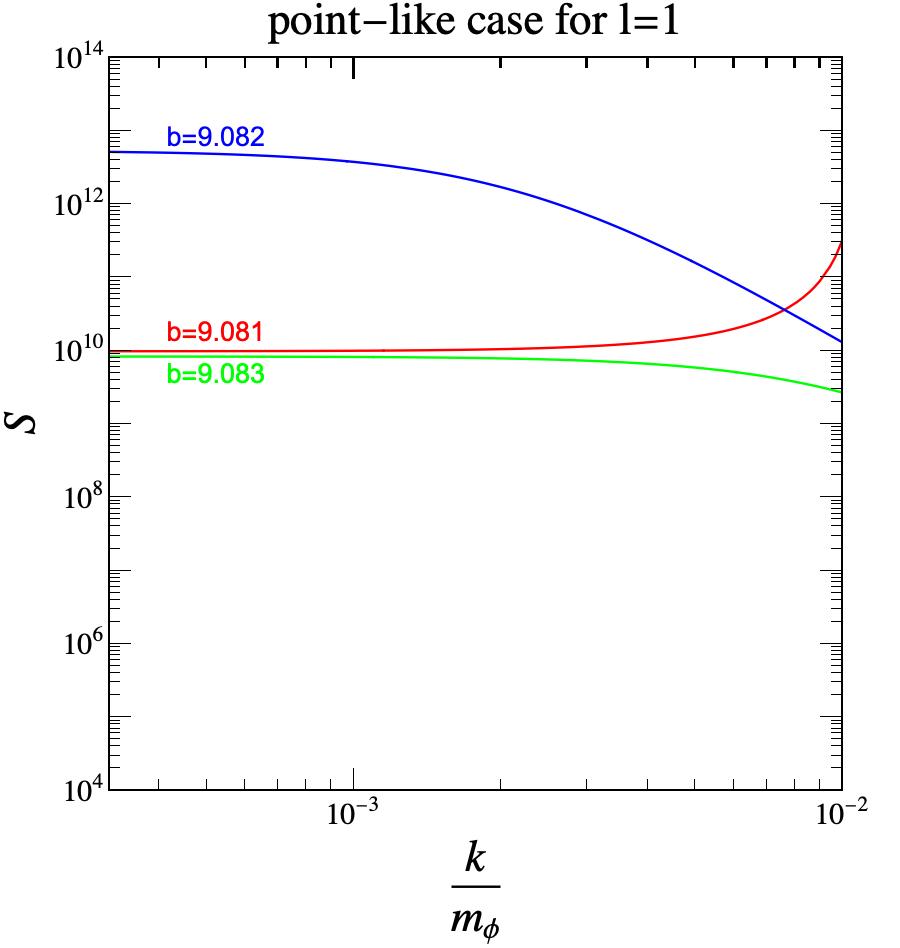}		
	\includegraphics[width=7.5cm]{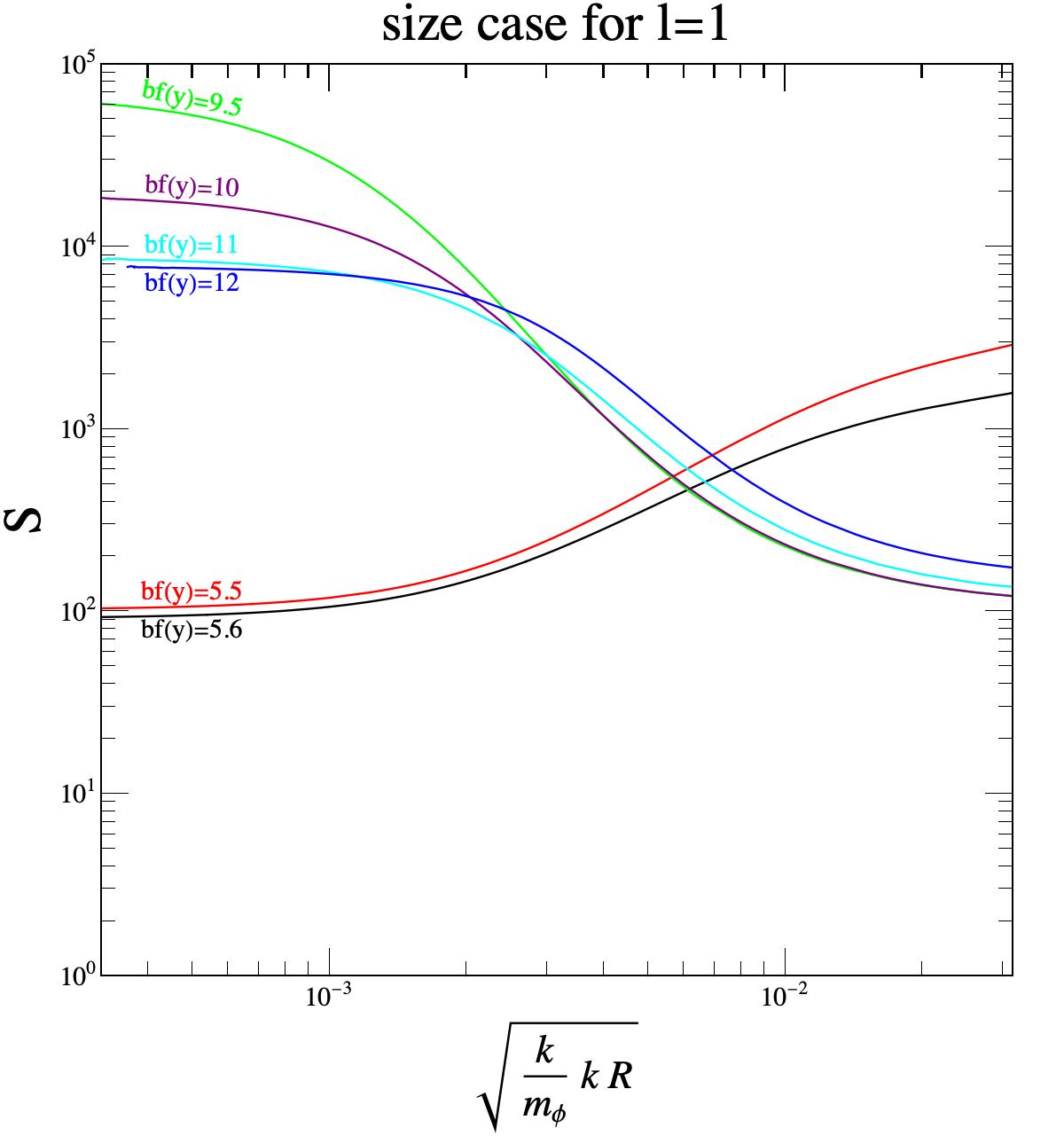}\\	
	\caption{Left panel: The p-wave sommerfeld enhancement factor $S$ as a function of $k/m_{\phi}$ with different values of parameter $b$ (
		near and at the first zero-energy resonance) for the point-like case;
		Right panel: The p-wave sommerfeld enhancement factor $S$ as a function of $\sqrt{(k/m_{\phi})(k R_{\chi})}$ with different values of parameter $bf(y)$ (near and at the first zero-energy resonance) for puffy dark matter.
	}\label{fig4}
\end{figure} 

To verify our conjecture numerically, we first examine the first zero-energy $p$-wave resonant state of point-like dark matter. It occurs at $b=9.082$ and is shown by the blue solid curve in the left panel of Fig.~\ref{fig4}. Its Sommerfeld enhancement factor is at least two orders of magnitude larger than those for $b=9.081$ and $b=9.083$. These two cases are shown by the red and green solid curves, respectively. For comparison, we also plot the first zero-energy $p$-wave resonant state of finite-size dark matter. The results are shown in the right panel of Fig.~\ref{fig4}. The solid curves correspond to $bf(y)=9.5$, 10, 11, and 12. Moreover, the upper-right panel of Fig.~\ref{fig3} shows that the first resonance appears near $bf(y)=10$. These results provide a mutual check. They demonstrate that the $p$-wave Sommerfeld resonance of finite-size dark matter without a specified internal structure is not a single peak. Instead, it extends over a finite width. Next, we study the $p$-wave resonant behavior of nugget dark matter with a specified internal structure.

\subsection{Nugget dark matter case}\label{sec3-2}
\begin{figure}[ht]
	\centering
	\includegraphics[width=8.4cm]{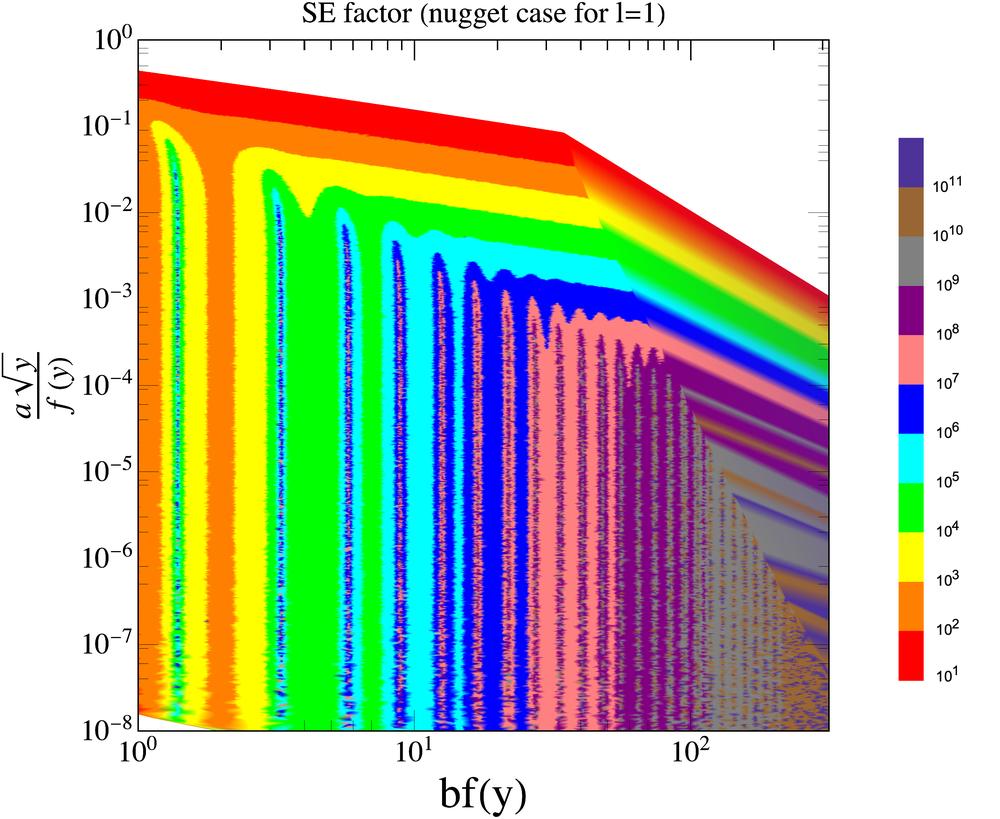}		
	\includegraphics[width=7.cm]{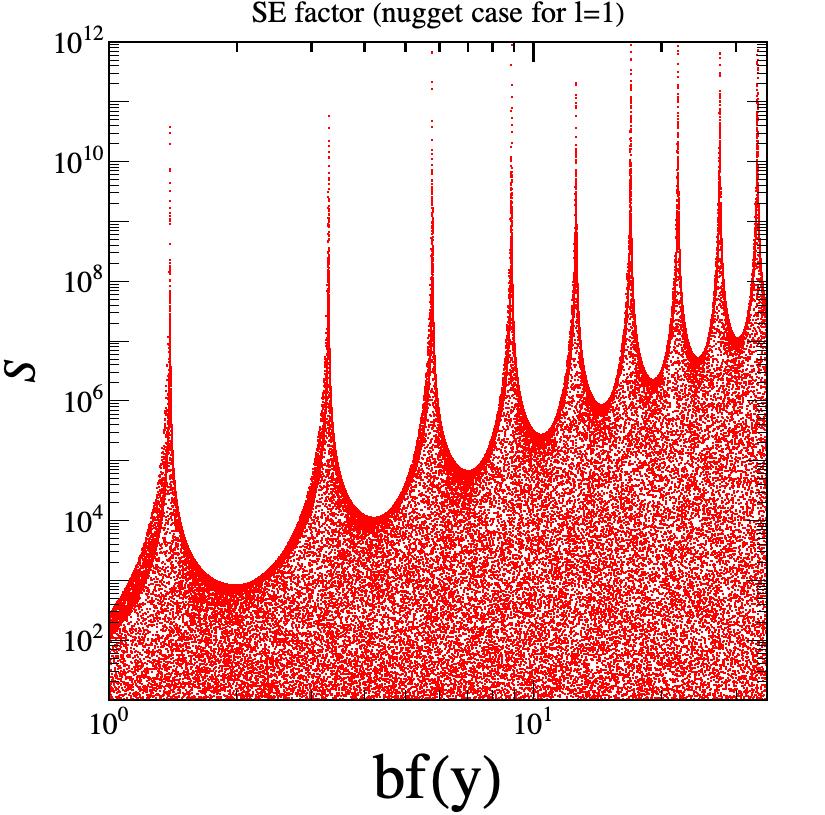}\\	
	\caption{Same as Fig.~\ref{fig3}, but the horizontal axis of the right panel is $b f(y)$ rather than $(b f(y))^4$.
	}
	\label{fig5}
\end{figure} 

As a finite-size dark matter candidate with an internal structure, we consider the nugget dark matter model. When a nugget contains a large number of constituents, its radius can become very large~\cite{Qi:2025dua,Bai:2024muo}. Consequently, the Sommerfeld enhancement becomes small unless a very light mediator is present. We therefore consider nugget dark matter with a small constituent number $N$. In this work, we choose $N=5$ for illustration. Following Ref.~\cite{Wise:2014jva}, we use the non-relativistic Fermi-gas approximation. The nugget radius can then be estimated as $R_{\chi}\simeq \left[\frac{81\pi^{2}}{4N g_{\rm dof}^{2}\alpha^{3}m_{\chi}^{3}}\right]^{1/3}$, where $g_{\rm dof}$ denotes the number of fermionic degrees of freedom. Under these conditions, we calculate the $p$-wave resonant behavior of nugget dark matter. The results are shown in the left panel of Fig.~\ref{fig5}. The $p$-wave resonances of nugget dark matter are similar to those of point-like dark matter. In both cases, the resonance positions and peaks are well defined. However, the resonance positions are different. This result can also be seen in the right panel of Fig.~\ref{fig5}. The finite resonance width disappears, and distinct resonance peaks emerge. Moreover, both panels of Fig.~\ref{fig5} show a smaller region of Sommerfeld enhancement. This region is smaller than that for point-like dark matter in the upper-left panel of Fig.~\ref{fig3}. Here, however, we consider only the case with $N=5$. We expect that the Sommerfeld-enhanced region and the resonance peaks may change with the constituent number. Therefore, future measurements of the resonance peak positions may allow us to infer the constituent number and radius of nugget dark matter.
\begin{figure}[ht]
	\centering
	\includegraphics[width=7.2cm]{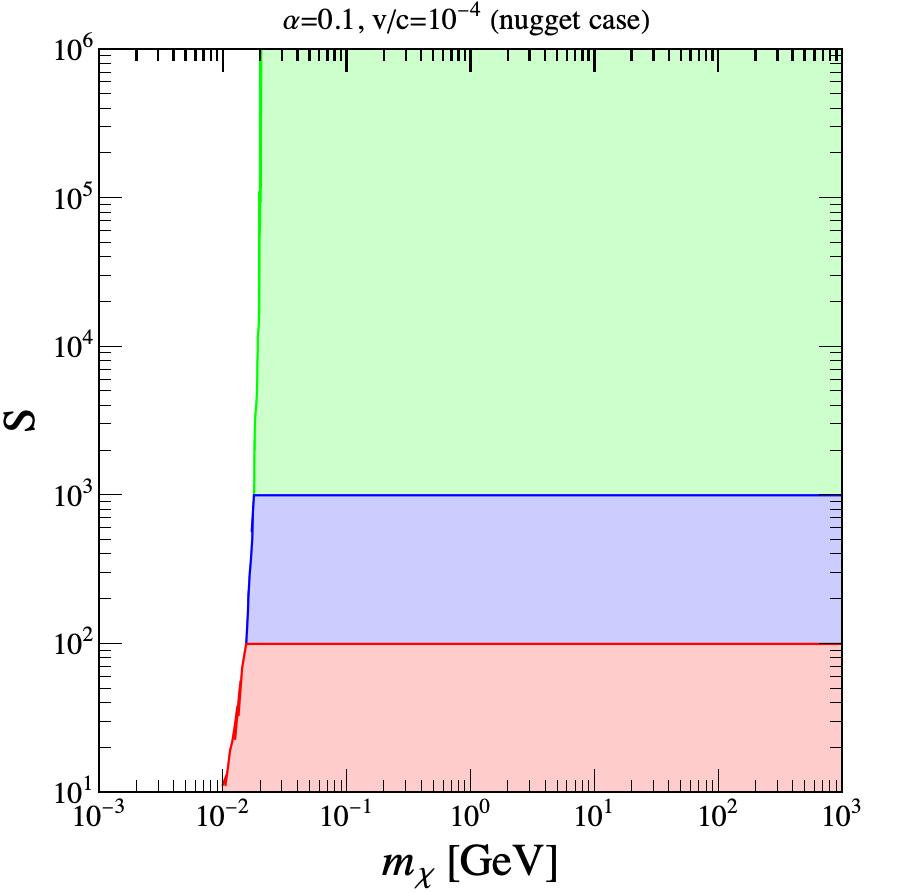}		
	\includegraphics[width=7.2cm]{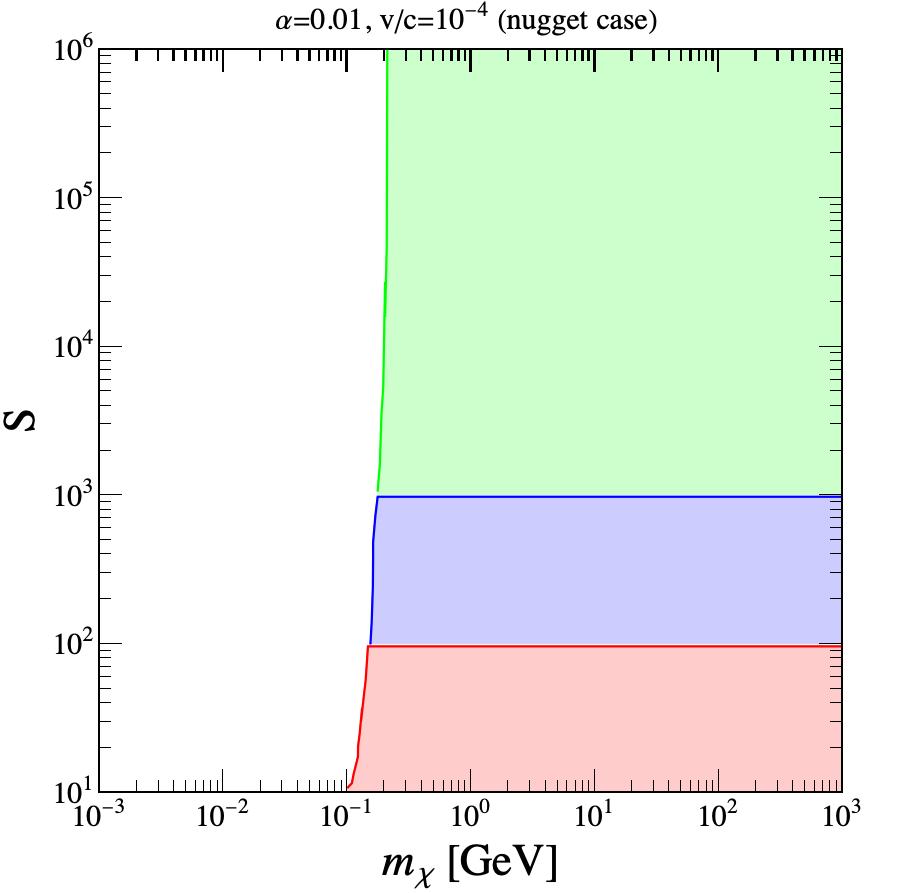}\\	
	\caption{The p-wave sommerfeld enhancement factor for nugget-type dark matter with $v/c=0.0001$ for different coupling constants: left panel ($\alpha=0.1$) and right panel ($\alpha=0.01$). Here the different color regions correspond to $S=10-100, 100-1000, 1000-10^6$, respectively. 
	}
	\label{fig6}
\end{figure} 

Finally, we present the viable parameter space for nugget dark matter with $N=5$. The results are shown in terms of the nugget mass $M_{\chi}=Nm_{\chi}$ and the Sommerfeld enhancement factor in Fig.~\ref{fig6}. Here, $m_{\chi}$ denotes the mass of a single dark matter constituent. It differs from the notation used in the previous sections. For example, $m_{\chi}$ on the horizontal axis of Fig.~\ref{fig1} denotes the total mass of finite-size dark matter. The left panel shows the case with $\alpha=0.1$. The corresponding lower bound on the constituent mass is approximately $m_{\chi}\sim10^{-2}~\mathrm{GeV}$. The right panel shows the result for $\alpha=0.01$. In this case, the allowed region requires $m_{\chi}>0.1~\mathrm{GeV}$.

In summary, the finite-size $p$-wave Sommerfeld resonances are weaker than those in the point-like case. Moreover, the resonances of finite-size dark matter have a finite width rather than a single peak. By contrast, nugget dark matter with a specified internal structure exhibits distinct $p$-wave resonance peaks. This behavior is similar to that of point-like dark matter. In the future, the resonance peak positions may allow us to determine the constituent number $N$ or the radius of nugget dark matter. The method used to analyze the $p$-wave resonances is similar to that used for the $s$-wave in our previous work~\cite{Xu:2026fsj}. However, we find that the $p$-wave contribution dominates the Sommerfeld enhancement of finite-size dark matter. Moreover, finite-size effects on the $p$-wave Sommerfeld enhancement may help reconcile the different annihilation cross-section requirements at different stages of cosmic evolution. Therefore, we consider it necessary to investigate the $p$-wave Sommerfeld resonances of finite-size dark matter.
\section{Conclusion}\label{sec4}

In this work, we analyzed the $p$-wave resonant behavior of finite-size dark matter. We first considered a finite-size dark matter object specified solely by its charge distribution. No specific internal structure was assumed. Our main results are as follows. First, as in the point-particle case, the $p$-wave contribution to the Sommerfeld enhancement is larger than the $s$-wave contribution. Second, finite-size effects reduce the magnitude of both the $s$-wave and $p$-wave Sommerfeld resonances. However, they leave the velocity dependence of the enhancement almost unchanged. Third, a finite-size object without a specified internal structure exhibits $p$-wave resonance regions with sharp boundaries but smeared internal structures. We then studied the $p$-wave resonances of nugget dark matter composed of a small number of constituents. After the internal structure is specified, the resonant behavior becomes similar to that of a point particle. In particular, the resonance peaks become narrower and better resolved. These results indicate that higher partial waves can provide the dominant contribution to the Sommerfeld resonances.

These changes in the resonant behavior may have important phenomenological implications. Previous analyses of the point-particle $p$-wave Sommerfeld enhancement imposed several observational and cosmological requirements. These include the dark matter relic density and the AMS-02 positron excess. They also include the Fermi-LAT constraints from dwarf spheroidal galaxies and the Planck energy-injection constraints during recombination. Together, these requirements strongly constrain the parameter space $(m_{\chi},m_{\phi},\alpha)$. Finite-size effects may relax some of these constraints. In the future, nonperturbative effects in finite-size dark matter models may help explain the different annihilation cross sections required at different stages of cosmic evolution. And nugget dark matter provides one possible example. Conversely, constraints on the annihilation cross section may help probe the internal structure of finite-size dark matter. For example, they may constrain the constituent number and radius of nugget dark matter.

\addcontentsline{toc}{section}{Acknowledgments}
\acknowledgments
 This work was supported by a Talent program from Chengdu Technological University under grant No. 2024RC031, by the National Natural Science Foundation of China (NNSFC) under grant No. 12335005 and by a PI Research Fund from Henan Normal University (5101029470335). 

\appendix
\section{The  potential function}\label{appa}
The interaction potential between two finite particles is 

\begin{align}\label{eqa}
V(r) & = \begin{cases}
- g(y) & r<2R_{\chi} \\
\hspace{5cm}\  & \ \\[-6.mm]
-\alpha\frac{e^{-m_{\phi}r}}{r} \times h\left(y\right)&  r>2R_{\chi},
\end{cases}
\end{align}
where
\begin{eqnarray} \label{inbol}
g(r,y)&=&-\alpha4\pi^{2}(\frac{3}{4\pi R_{\chi}^{3}})^{2}\frac{(1+y)e^{-y}}{m_{\phi^{3}}}\frac{1}{2m_{\phi}^{3}} \nonumber \\
&& \times\left(\frac{2e^{-m_{\phi}(R_{\chi}+r)}(-1+e^{y})(e^{2y}(-1+y)+e^{y}(1+y))}{r}\right.\nonumber \\
&& \Bigg.+m_{\phi}(-e^{y}(2+m_{\phi}(r-2R_{\chi}))+e^{-y}(2-m_{\phi}r+2y))\Bigg)\nonumber \\
&& +\alpha4\pi^{2}\left(\frac{3}{4\pi R_{\chi}^{3}}\right)^{2}\frac{1}{m_{\phi}}\left(\frac{e^{-y}}{m_{\phi}^{2}}-\frac{e^{y}}{m_{\phi}^{2}}+\frac{R_{\chi}e^{-y}}{m_{\phi}}+\frac{R_{\chi}e^{y}}{m_{\phi}}\right) \nonumber \\
&& \times\frac{e^{-m_{\phi}(R_{\chi}+r)}(2+2y-e^{y}(2+m_{\phi}r(-2+m_{\phi}(r-2R_{\chi}))+2y))}{2m_{\phi}^{3}r}\nonumber  \\
&& +\alpha4\pi^{2}\left(\frac{3}{4\pi R_{\chi}^{3}}\right)^{2}\frac{1}{m_{\phi}^2}\left( \frac{1}{12}(r-2R_{\chi})(r+4R_{\chi})\right), \\
h\left(y\right)&=&4\pi\left(\frac{3}{4\pi R_{\chi}^{3}}\right)^{2}\frac{\pi}{m_{\phi}^{2}}\left(\frac{e^{-y}}{m_{\phi}^{2}}-\frac{e^{y}}{m_{\phi}^{2}}+\frac{R_{\chi}e^{-y}}{m_{\phi}}+\frac{R_{\chi}e^{y}}{m_{\phi}}\right)^{2}.
\end{eqnarray}
\addcontentsline{toc}{section}{References}
\bibliographystyle{JHEP}
\bibliography{note}

@article{Bai:2024muo,
	author = "Bai, Yang and Korwar, Mrunal",
	title = "{Radioactivity of quark nuggets}",
	eprint = "2409.16487",
	archivePrefix = "arXiv",
	primaryClass = "hep-ph",
	reportNumber = "N3AS-24-034",
	doi = "10.1007/JHEP06(2025)059",
	journal = "JHEP",
	volume = "06",
	pages = "059",
	year = "2025"
}

@article{Qi:2025dua,
	author = "Qi, Haoyang and Qi, Hao-Yang and Xu, Renxin and Xu, Ren-Xin",
	title = "{To Detect Strong Nugget with an Acoustic Array}",
	eprint = "2503.20597",
	archivePrefix = "arXiv",
	primaryClass = "astro-ph.IM",
	doi = "10.1088/1674-4527/ade953",
	journal = "Res. Astron. Astrophys.",
	volume = "25",
	number = "9",
	pages = "095010",
	year = "2025"
}

@article{Hartl:2025mtb,
	author = "Hartl, Odelia V. and Vienneau, Evan and Batteas, Evan and Evans, Addy J. and Bozorgnia, Nassim and Strigari, Louis E.",
	title = "{Enhancements in velocity-dependent dark matter annihilation in Galactic subhalos}",
	eprint = "2509.05519",
	archivePrefix = "arXiv",
	primaryClass = "astro-ph.HE",
	doi = "10.1088/1475-7516/2026/02/024",
	journal = "JCAP",
	volume = "02",
	pages = "024",
	year = "2026"
}

@article{Pradhan:2026maz,
	author = "Pradhan, Dipankar and Mondal, Niloy and Sarkar, Abhik and Ghosh, Anupam and Sharma, Shashwat and Thomas Arun, Mathew and Barman, Basabendu",
	title = "{From WIMP to FIMP during reheating: collider vs non-collider probes for p-wave annihilation}",
	eprint = "2605.27521",
	archivePrefix = "arXiv",
	primaryClass = "hep-ph",
	month = "5",
	year = "2026"
}

@article{Arkani-Hamed:2008hhe,
	author = "Arkani-Hamed, Nima and Finkbeiner, Douglas P. and Slatyer, Tracy R. and Weiner, Neal",
	title = "{A Theory of Dark Matter}",
	eprint = "0810.0713",
	archivePrefix = "arXiv",
	primaryClass = "hep-ph",
	doi = "10.1103/PhysRevD.79.015014",
	journal = "Phys. Rev. D",
	volume = "79",
	pages = "015014",
	year = "2009"
}

@article{Feng:2010zp,
	author = "Feng, Jonathan L. and Kaplinghat, Manoj and Yu, Hai-Bo",
	title = "{Sommerfeld Enhancements for Thermal Relic Dark Matter}",
	eprint = "1005.4678",
	archivePrefix = "arXiv",
	primaryClass = "hep-ph",
	reportNumber = "UCI-TR-2010-06",
	doi = "10.1103/PhysRevD.82.083525",
	journal = "Phys. Rev. D",
	volume = "82",
	pages = "083525",
	year = "2010"
}

@article{Das:2016ced,
	author = "Das, Anirban and Dasgupta, Basudeb",
	title = "{Selection Rule for Enhanced Dark Matter Annihilation}",
	eprint = "1611.04606",
	archivePrefix = "arXiv",
	primaryClass = "hep-ph",
	reportNumber = "TIFR-TH-16-43",
	doi = "10.1103/PhysRevLett.118.251101",
	journal = "Phys. Rev. Lett.",
	volume = "118",
	number = "25",
	pages = "251101",
	year = "2017"
}

@article{Beneke:2024iev,
	author = "Beneke, Martin and Binder, Tobias and De Ros, Lorenzo and Garny, Mathias",
	title = "{Enhancement of p-wave dark matter annihilation by quasi-bound states}",
	eprint = "2403.07108",
	archivePrefix = "arXiv",
	primaryClass = "hep-ph",
	reportNumber = "TUM-HEP-1500/24",
	doi = "10.1007/JHEP06(2024)207",
	journal = "JHEP",
	volume = "06",
	pages = "207",
	year = "2024"
}

@article{Wang:2025tdx,
	author = "Wang, Yu-Ning and Duan, Xin-Chen and Tang, Tian-Peng and Wang, Ziwei and Tsai, Yue-Lin Sming",
	title = "{Exploring sub-GeV dark matter via s-wave, p-wave, and resonance annihilation with CMB data}",
	eprint = "2502.18263",
	archivePrefix = "arXiv",
	primaryClass = "hep-ph",
	doi = "10.1088/1475-7516/2025/08/059",
	journal = "JCAP",
	volume = "08",
	pages = "059",
	year = "2025"
}

@article{Iengo:2009ni,
	author = "Iengo, Roberto",
	title = "{Sommerfeld enhancement: General results from field theory diagrams}",
	eprint = "0902.0688",
	archivePrefix = "arXiv",
	primaryClass = "hep-ph",
	doi = "10.1088/1126-6708/2009/05/024",
	journal = "JHEP",
	volume = "05",
	pages = "024",
	year = "2009"
}

@article{Kamada:2023iol,
	author = "Kamada, Ayuki and Kuwahara, Takumi and Patel, Ami",
	title = "{Quantum theory of dark matter scattering}",
	eprint = "2303.17961",
	archivePrefix = "arXiv",
	primaryClass = "hep-ph",
	doi = "10.1007/JHEP11(2023)105",
	journal = "JHEP",
	volume = "11",
	pages = "105",
	year = "2023"
}

@article{Parikh:2024mwa,
	author = "Parikh, Aditya and Sato, Ryosuke and Slatyer, Tracy R.",
	title = "{Regulating Sommerfeld resonances for multi-state systems and higher partial waves}",
	eprint = "2410.18168",
	archivePrefix = "arXiv",
	primaryClass = "hep-ph",
	reportNumber = "MIT-CTP/5790, OU-HET-1243",
	doi = "10.1007/JHEP12(2025)025",
	journal = "JHEP",
	volume = "12",
	pages = "025",
	year = "2025"
}

@article{An:2012uu,
	author = "An, Haipeng and Pospelov, Maxim",
	title = "{Dark Resonance}",
	eprint = "1206.2371",
	archivePrefix = "arXiv",
	primaryClass = "hep-ph",
	reportNumber = "PI-PARTPHYS-281",
	doi = "10.1088/1475-7516/2012/11/061",
	journal = "JCAP",
	volume = "11",
	pages = "061",
	year = "2012"
}

@article{Watanabe:2025kgw,
	author = "Watanabe, Yuki",
	title = "{Unitarization of the Sommerfeld enhancement through the renormalization group}",
	eprint = "2508.09511",
	archivePrefix = "arXiv",
	primaryClass = "hep-ph",
	doi = "10.1007/JHEP01(2026)131",
	journal = "JHEP",
	volume = "01",
	pages = "131",
	year = "2026"
}

@article{Sommerfeld:1931qaf,
	author = "Sommerfeld, A.",
	title = {{{\"U}ber die Beugung und Bremsung der Elektronen}},
	doi = "10.1002/andp.19314030302",
	journal = "Annalen Phys.",
	volume = "403",
	number = "3",
	pages = "257--330",
	year = "1931"
}

@article{Daylan:2014rsa,
	author = "Daylan, Tansu and Finkbeiner, Douglas P. and Hooper, Dan and Linden, Tim and Portillo, Stephen K. N. and Rodd, Nicholas L. and Slatyer, Tracy R.",
	title = "{The Characterization of the Gamma-Ray Signal from the Central Milky Way: A Case for Annihilating Dark Matter}",
	eprint = "1402.6703",
	archivePrefix = "arXiv",
	primaryClass = "astro-ph.HE",
	reportNumber = "FERMILAB-PUB-14-032-A, MIT-CTP-4533",
	doi = "10.1016/j.dark.2015.12.005",
	journal = "Phys. Dark Univ.",
	volume = "12",
	pages = "1--23",
	year = "2016"
}

@article{Prantzos:2010wi,
	author = "Prantzos, N. and others",
	title = "{The 511 keV emission from positron annihilation in the Galaxy}",
	eprint = "1009.4620",
	archivePrefix = "arXiv",
	primaryClass = "astro-ph.HE",
	doi = "10.1103/RevModPhys.83.1001",
	journal = "Rev. Mod. Phys.",
	volume = "83",
	pages = "1001--1056",
	year = "2011"
}

@article{Fermi-LAT:2025gei,
	author = "Abdollahi, S. and others",
	collaboration = "Fermi-LAT, HAWC, H.E.S.S., MAGIC, VERITAS",
	title = "{Combined dark matter search towards dwarf spheroidal galaxies with Fermi-LAT, HAWC, H.E.S.S., MAGIC, and VERITAS}",
	eprint = "2508.20229",
	archivePrefix = "arXiv",
	primaryClass = "astro-ph.HE",
	doi = "10.1088/1475-7516/2026/03/035",
	journal = "JCAP",
	volume = "03",
	pages = "035",
	year = "2026"
}

@article{Planck:2018vyg,
	author = "Aghanim, N. and others",
	collaboration = "Planck",
	title = "{Planck 2018 results. VI. Cosmological parameters}",
	eprint = "1807.06209",
	archivePrefix = "arXiv",
	primaryClass = "astro-ph.CO",
	doi = "10.1051/0004-6361/201833910",
	journal = "Astron. Astrophys.",
	volume = "641",
	pages = "A6",
	year = "2020",
	note = "[Erratum: Astron.Astrophys. 652, C4 (2021)]"
}

@article{Masi:2015fca,
	author = "Masi, Nicol{\`o} and Ballardini, Mario",
	title = "{A conservative assessment of the current constraints on dark matter annihilation from Cosmic Rays and CMB observations}",
	eprint = "1509.00058",
	archivePrefix = "arXiv",
	primaryClass = "astro-ph.CO",
	doi = "10.1142/S0218271817500419",
	journal = "Int. J. Mod. Phys. D",
	volume = "26",
	number = "06",
	pages = "1750041",
	year = "2016"
}

@article{Ding:2021zzg,
	author = "Ding, Yu-Chen and Ku, Yu-Lin and Wei, Chun-Cheng and Zhou, Yu-Feng",
	title = "{Consistent explanation for the cosmic-ray positron excess in p-wave Sommerfeld-enhanced dark matter annihilation}",
	eprint = "2104.14881",
	archivePrefix = "arXiv",
	primaryClass = "hep-ph",
	doi = "10.1088/1475-7516/2021/09/005",
	journal = "JCAP",
	volume = "09",
	pages = "005",
	year = "2021"
}

@article{Xu:2025xaq,
	author = "Xu, Wu-Long and Yang, Jin Min and Zhao, Jun",
	title = "{Scattering of non-relativistic finite-size particles and puffy dark matter direct detection}",
	eprint = "2510.10641",
	archivePrefix = "arXiv",
	primaryClass = "hep-ph",
	doi = "10.1007/JHEP04(2026)006",
	journal = "JHEP",
	volume = "04",
	pages = "006",
	year = "2026"
}

@article{Xu:2020qjk,
	author = "Xu, Xingchen and Farrar, Glennys R.",
	title = "{Resonant scattering between dark matter and baryons: Revised direct detection and CMB limits}",
	eprint = "2101.00142",
	archivePrefix = "arXiv",
	primaryClass = "hep-ph",
	doi = "10.1103/PhysRevD.107.095028",
	journal = "Phys. Rev. D",
	volume = "107",
	number = "9",
	pages = "095028",
	year = "2023"
}

@article{Wang:2021tjf,
	author = "Wang, Wenyu and Xu, Wu-Long and Zhu, Bin",
	title = "{Realistic scattering of puffy dark matter}",
	eprint = "2108.07030",
	archivePrefix = "arXiv",
	primaryClass = "hep-ph",
	doi = "10.1103/PhysRevD.105.075013",
	journal = "Phys. Rev. D",
	volume = "105",
	number = "7",
	pages = "075013",
	year = "2022"
}

@article{Wang:2023wrx,
	author = "Wang, Wenyu and Xu, Wu-Long and Yang, Jin Min and Zhu, Rui",
	title = "{Direct detection of cosmic ray-boosted puffy dark matter}",
	eprint = "2305.12668",
	archivePrefix = "arXiv",
	primaryClass = "hep-ph",
	doi = "10.1016/j.nuclphysb.2023.116348",
	journal = "Nucl. Phys. B",
	volume = "995",
	pages = "116348",
	year = "2023"
}

@article{Wang:2023wbw,
	author = "Wang, Wenyu and Xu, Wu-Long and Yang, Jin Min and Zhu, Bin and Zhu, Rui",
	title = "{Sommerfeld enhancement for puffy self-interacting dark matter}",
	eprint = "2308.02170",
	archivePrefix = "arXiv",
	primaryClass = "hep-ph",
	doi = "10.1007/JHEP01(2024)114",
	journal = "JHEP",
	volume = "01",
	pages = "114",
	year = "2024"
}

@article{Xu:2026fsj,
	author = "Xu, Wu-Long and Yang, Jin Min and Yang, Wen-Na",
	title = "{Size Dependence of the Sommerfeld Enhancement for Puffy Dark Matter}",
	eprint = "2606.29710",
	archivePrefix = "arXiv",
	primaryClass = "hep-ph",
	month = "6",
	year = "2026"
}

@article{Chu:2018faw,
	author = "Chu, Xiaoyong and Garcia-Cely, Camilo and Murayama, Hitoshi",
	title = "{Finite-size dark matter and its effect on small-scale structure}",
	eprint = "1901.00075",
	archivePrefix = "arXiv",
	primaryClass = "hep-ph",
	reportNumber = "DESY-18-225, IPMU18-0207",
	doi = "10.1103/PhysRevLett.124.041101",
	journal = "Phys. Rev. Lett.",
	volume = "124",
	number = "4",
	pages = "041101",
	year = "2020"
}

@article{Bringmann:2016din,
	author = "Bringmann, Torsten and Kahlhoefer, Felix and Schmidt-Hoberg, Kai and Walia, Parampreet",
	title = "{Strong constraints on self-interacting dark matter with light mediators}",
	eprint = "1612.00845",
	archivePrefix = "arXiv",
	primaryClass = "hep-ph",
	reportNumber = "DESY-16-226",
	doi = "10.1103/PhysRevLett.118.141802",
	journal = "Phys. Rev. Lett.",
	volume = "118",
	number = "14",
	pages = "141802",
	year = "2017"
}

@article{Belli:2026woj,
	author = "Belli, Pierluigi and Bernabei, Rita and Beylin, Vitaly and Bikbaev, Timur and Kharakhashyan, Artem and Khlopov, Maxim and Korchagin, Vladimir and Mayorov, Andrey and Sopin, Danila",
	title = "{A Dark Atom Scenario for Direct Dark Matter Investigation}",
	doi = "10.3390/universe12040116",
	journal = "Universe",
	volume = "12",
	number = "4",
	pages = "116",
	year = "2026"
}

@article{Grabowska:2018lnd,
	author = "Grabowska, Dorota M. and Melia, Tom and Rajendran, Surjeet",
	title = "{Detecting Dark Blobs}",
	eprint = "1807.03788",
	archivePrefix = "arXiv",
	primaryClass = "hep-ph",
	doi = "10.1103/PhysRevD.98.115020",
	journal = "Phys. Rev. D",
	volume = "98",
	number = "11",
	pages = "115020",
	year = "2018"
}

@article{Wise:2014ola,
	author = "Wise, Mark B. and Zhang, Yue",
	title = "{Yukawa Bound States of a Large Number of Fermions}",
	eprint = "1411.1772",
	archivePrefix = "arXiv",
	primaryClass = "hep-ph",
	reportNumber = "CALT-TH-2014-160",
	doi = "10.1007/JHEP02(2015)023",
	journal = "JHEP",
	volume = "02",
	pages = "023",
	year = "2015",
	note = "[Erratum: JHEP 10, 165 (2015)]"
}

@article{Wise:2014jva,
	author = "Wise, Mark B. and Zhang, Yue",
	title = "{Stable Bound States of Asymmetric Dark Matter}",
	eprint = "1407.4121",
	archivePrefix = "arXiv",
	primaryClass = "hep-ph",
	reportNumber = "CALT-TH-2014-145",
	doi = "10.1103/PhysRevD.90.055030",
	journal = "Phys. Rev. D",
	volume = "90",
	number = "5",
	pages = "055030",
	year = "2014",
	note = "[Erratum: Phys.Rev.D 91, 039907 (2015)]"
}

@article{Gresham:2017cvl,
	author = "Gresham, Moira I. and Lou, Hou Keong and Zurek, Kathryn M.",
	title = "{Early Universe synthesis of asymmetric dark matter nuggets}",
	eprint = "1707.02316",
	archivePrefix = "arXiv",
	primaryClass = "hep-ph",
	doi = "10.1103/PhysRevD.97.036003",
	journal = "Phys. Rev. D",
	volume = "97",
	number = "3",
	pages = "036003",
	year = "2018"
}

@article{Gresham:2017zqi,
	author = "Gresham, Moira I. and Lou, Hou Keong and Zurek, Kathryn M.",
	title = "{Nuclear Structure of Bound States of Asymmetric Dark Matter}",
	eprint = "1707.02313",
	archivePrefix = "arXiv",
	primaryClass = "hep-ph",
	doi = "10.1103/PhysRevD.96.096012",
	journal = "Phys. Rev. D",
	volume = "96",
	number = "9",
	pages = "096012",
	year = "2017"
}

@article{Bai:2018dxf,
	author = "Bai, Yang and Long, Andrew J. and Lu, Sida",
	title = "{Dark Quark Nuggets}",
	eprint = "1810.04360",
	archivePrefix = "arXiv",
	primaryClass = "hep-ph",
	reportNumber = "FERMILAB-PUB-18-600-T",
	doi = "10.1103/PhysRevD.99.055047",
	journal = "Phys. Rev. D",
	volume = "99",
	number = "5",
	pages = "055047",
	year = "2019"
}

@article{Monteiro:2020wcb,
	author = "Monteiro, Fernando and Afek, Gadi and Carney, Daniel and Krnjaic, Gordan and Wang, Jiaxiang and Moore, David C.",
	title = "{Search for composite dark matter with optically levitated sensors}",
	eprint = "2007.12067",
	archivePrefix = "arXiv",
	primaryClass = "hep-ex",
	reportNumber = "FERMILAB-PUB-20-282-AE-T",
	doi = "10.1103/PhysRevLett.125.181102",
	journal = "Phys. Rev. Lett.",
	volume = "125",
	number = "18",
	pages = "181102",
	year = "2020"
}

@article{Carenza:2024qaq,
	author = "Carenza, Pierluca and Pasechnik, Roman and Wang, Zhi-Wei",
	title = "{Composite Heavy Axionlike Dark Matter}",
	eprint = "2408.14245",
	archivePrefix = "arXiv",
	primaryClass = "hep-ph",
	doi = "10.1103/49pk-z8c8",
	journal = "Phys. Rev. Lett.",
	volume = "135",
	number = "2",
	pages = "021001",
	year = "2025"
}

@article{Sannino:2026hkc,
	author = "Sannino, Francesco and Turner, Jessica",
	title = "{Interpreting the LZ 248 keV Event using Dark QCD}",
	eprint = "2609.24988",
	archivePrefix = "arXiv",
	primaryClass = "hep-ph",
	reportNumber = "IPPP/26/65",
	month = "9",
	year = "2026"
}

@article{Cline:2021itd,
	author = "Cline, James M.",
	title = "{Dark atoms and composite dark matter}",
	eprint = "2108.10314",
	archivePrefix = "arXiv",
	primaryClass = "hep-ph",
	doi = "10.21468/SciPostPhysLectNotes.52",
	journal = "SciPost Phys. Lect. Notes",
	volume = "52",
	pages = "1",
	year = "2022"
}

@article{Lin:2019uvt,
	author = "Lin, Tongyan",
	title = "{Dark matter models and direct detection}",
	eprint = "1904.07915",
	archivePrefix = "arXiv",
	primaryClass = "hep-ph",
	doi = "10.22323/1.333.0009",
	journal = "PoS",
	volume = "333",
	pages = "009",
	year = "2019"
}

@article{XENON:2025vwd,
	author = "Aprile, E. and others",
	collaboration = "XENON",
	title = "{WIMP Dark Matter Search Using a 3.1 Tonne-Year Exposure of the XENONnT Experiment}",
	eprint = "2502.18005",
	archivePrefix = "arXiv",
	primaryClass = "hep-ex",
	doi = "10.1103/msw4-t342",
	journal = "Phys. Rev. Lett.",
	volume = "135",
	number = "22",
	pages = "221003",
	year = "2025"
}

@article{Jungman:1995df,
	author = "Jungman, Gerard and Kamionkowski, Marc and Griest, Kim",
	title = "{Supersymmetric dark matter}",
	eprint = "hep-ph/9506380",
	archivePrefix = "arXiv",
	reportNumber = "SU-4240-605, UCSD-PTH-95-02, IASSNS-HEP-95-14, CU-TP-677",
	doi = "10.1016/0370-1573(95)00058-5",
	journal = "Phys. Rept.",
	volume = "267",
	pages = "195--373",
	year = "1996"
}

@article{Buchmueller:2017qhf,
	author = "Buchmueller, Oliver and Doglioni, Caterina and Wang, Lian Tao",
	title = "{Search for dark matter at colliders}",
	eprint = "1912.12739",
	archivePrefix = "arXiv",
	primaryClass = "hep-ex",
	doi = "10.1038/nphys4054",
	journal = "Nature Phys.",
	volume = "13",
	number = "3",
	pages = "217--223",
	year = "2017"
}

@article{Clowe:2006eq,
	author = "Clowe, Douglas and Bradac, Marusa and Gonzalez, Anthony H. and Markevitch, Maxim and Randall, Scott W. and Jones, Christine and Zaritsky, Dennis",
	title = "{A direct empirical proof of the existence of dark matter}",
	eprint = "astro-ph/0608407",
	archivePrefix = "arXiv",
	reportNumber = "SLAC-PUB-12078",
	doi = "10.1086/508162",
	journal = "Astrophys. J. Lett.",
	volume = "648",
	pages = "L109--L113",
	year = "2006"
}

@article{Cirelli:2024ssz,
	author = "Cirelli, Marco and Strumia, Alessandro and Zupan, Jure",
	title = "{Dark Matter}",
	eprint = "2406.01705",
	archivePrefix = "arXiv",
	primaryClass = "hep-ph",
	doi = "10.21468/SciPostPhysRev.1",
	month = "6",
	year = "2024"
}

@article{Wang:2023xii,
    author = "Wang, Wenyu and Xu, Wu-Long and Yang, Jin Min and Zhu, Bin",
    title = "{Revisiting puffy dark matter with novel insights: partial wave analysis}",
    eprint = "2303.11058",
    archivePrefix = "arXiv",
    primaryClass = "hep-ph",
    doi = "10.1007/JHEP06(2023)103",
    journal = "JHEP",
    volume = "06",
    pages = "103",
    year = "2023"
}

\end{document}